\documentclass[12pt]{article}

\usepackage{a4wide}
\usepackage[pdftex,usenames,dvipsnames]{color}
\usepackage{graphics}
\usepackage{amsfonts}
\usepackage{amssymb}
\usepackage{accents}

\newcommand{\nit}{\noindent}

\newcommand{\np}{\newpage}
\newcommand{\dsp}{\displaystyle}
\newcommand{\vs}[1]{\vspace{#1 ex}}
\newcommand{\hs}[1]{\hspace{#1 em}}
\newcommand{\bfr}{\begin{flushright}}
\newcommand{\efr}{\end{flushright}}
\newcommand{\bc}{\begin{center}}
\newcommand{\ec}{\end{center}}
\newcommand{\ben}{\begin{enumerate}}
\newcommand{\een}{\end{enumerate}}

\newcommand{\be}{\begin{equation}}
\newcommand{\ee}{\end{equation}}
\newcommand{\ba}{\begin{array}}
\newcommand{\ea}{\end{array}}
\newcommand{\ct}{\cite}
\newcommand{\bit}{\bibitem}

\newcommand{\gam}{\gamma}
\newcommand{\del}{\delta}

\newcommand{\lb}{\lambda}
\newcommand{\sg}{\sigma}

\newcommand{\vf}{\varphi}

\newcommand{\Del}{\Delta}

\newcommand{\bfxi}{\mbox{{\boldmath $\xi$}}}

\newcommand{\bfa}{\bold{a}}
\newcommand{\bfb}{\bold{b}}

\newcommand{\bfn}{{\bf n}}
\newcommand{\bfp}{\bold{p}}

\newcommand{\bfer}{\bold{r}}

\newcommand{\bfu}{\bold{u}}
\newcommand{\bfv}{\bold{v}}

\newcommand{\bfx}{\bold{x}}

\newcommand{\bfF}{\bold {F}}

\newcommand{\bfL}{\bold {L}}

\newcommand{\bfP}{\bold {P}}

\newcommand{\bfX}{\bold {X}}

\newcommand{\lh}{\left(}
\newcommand{\rh}{\right)}

\newcommand{\nb}{\nabla}

\newcommand{\der}{\partial}

\begin{document}

\pagestyle{plain}
\pagenumbering{arabic}

\nit
\bc
{\bf A Huygens-Leibniz-Lange framework for classical mechanics}
\vs{2}

Jan-Willem van Holten 
\vs{1}

Lorentz Institute, Leiden University, \\
and \\
Nikhef, Amsterdam, NL 
\vs{2}

\today
\ec
\vs{3}

\nit
{\small
{\bf Abstract }\\
I discuss the physical basis of classical mechanics, such as expressed commonly using 
the framework of Newton's {\em Principia}. Newton's formulation of the laws of motion is 
seen to have quite a few ambiguities and shortcomings. Therefore I offer an alternative set 
of laws, based in particular on ideas of his contemporaries Huygens and Leibniz with a crucial 
addition by Ludwig Lange, which avoids the problems with Newton's formulation. It is shown 
that from these laws of motion all the usual results of classical mechanics, as it concerns the 
motion of idealized point masses, can be rederived. The application of these principles to 
relativistic point particles is discussed.  }
\vs{2}

\nit
{\bf Introduction}
\vs{1}

\nit
There exist a great many different formulations of the theory of classical mechanics. It has been 
phrased in terms of systems of first or second order differential equations, variational principles, 
phase-space methods and Hamilton-Jacobi equations, to name the most common ones. 
These various approaches provide as many different lines of attack to solve mathematical 
problems arising in the theory; indeed, they have in common that they mostly concern the 
mathematical formulation of the theory, without touching the underlying physical principles, for 
which they usually refer to the work of Newton \ct{newton:1686}, and sometimes 
Galileo \ct{galileo:1632} and Huygens \ct{huygens:1673}.

Surprisingly --with the exception of the development of the theory of Relativity \ct{einstein:1905}-- 
these physical principles underlying the theory have been getting much less attention. One gets the 
impression that after the work of Ernst Mach in the second half of the nineteenth century 
\ct{mach:1883} the majority of authors considered the physical principles on which Newton's 
laws of motion were based as understood and in no more need of clarification. 

That this is not the case, and that Newton's formulation of the basic laws of motion are sometimes
ambiguous, incomplete or even inconsistent, can be seen already in the reactions of his 
contemporaries. Heated debates about the status of concepts such as space, time, the vacuum 
and force were commonplace. Metaphysical and theological interpretations were offered and 
rejected; examples of these can be found among others in the essays published by learned 
societies, like the Royal Society in England, the Paris Acad\'{e}mie Royale des Sciences and 
the Acad\'{e}mie Royale des Sciences et Belles-Lettres in Berlin, especially in the first half of 
the eighteenth century. The discussions also reached the public at large in such semi-popular 
accounts as those of van Musschenbroek in the Netherlands \ct{musschenbroek:1736} and du 
Ch\^{a}telet \ct{duchatelet:1740} in France. An excellent and broad historical review of the issues 
involved is found in Jammer's lectures \ct{jammer:1957}. A more recent in-depth study is that of 
Boudri \ct{boudri:1994}.

In the following I present an analysis of the most pressing issues concerning the laws of 
motion as presented by Newton. This is followed by a reconstruction of the basics using observable 
kinematic quantities which can be measured directly in terms of properly defined distances and 
time intervals. Most of the fundamental ideas can in fact be found in the earlier work of Galileo 
and especially Huygens and Leibniz. Another absolutely crucial concept needed is that of inertial 
frames, introduced in 1885 by Ludwig Lange \ct{lange:1885}.
\vs{2}

\nit
{\bf 1.\ Motion in classical mechanics}
\vs{1}

\nit
Ever since its publication Newton's {\em Principia Mathematica} roused intense debates about his 
formulation of the laws of motion. For completeness, let me recall these well-known laws in the 
English translation of Andrew Motte (1729) \ct{newton:1686}: 
\vs{1}

\nit
1.\ {\em Every body continues in its state of rest, or of uniform motion in a right line, unless it is compelled
to change that state by forces impressed on it. }
\vs{1}

\nit
2.\ {\em The change of motion is proportional to the motive force impressed; and it is made in the 
direction of the right line in which that force is impressed. }
\vs{1}

\nit
3.\ {\em To every action there is always apposed an equal reaction: or, the mutual actions of two bodies upon 
each other are always equal, and directed to contrary parts. }
\vs{1}

\nit
The first law is the principle of inertia, first given in this form by Huygens \ct{huygens:1669,huygens:1703}.
In the second law {\em change of motion} refers to the momentum, the product of mass and velocity: 
$\bfp = m \bfv$ as a vector, taking direction into account; then its content can be expressed in modern 
mathematics by the equation 
\be
\bfF = \frac{d\bfp}{dt},
\label{1.1}
\ee
which for bodies of fixed mass $m$ reduces to the familiar form 
\be
\bfF = m \bfa.
\label{1.2}
\ee
Finally, the third law implies that {\em force} is not an external agent, but acts mutually between bodies 
in such a way that the action of the first on the second is equally large as that of the second on the first,
but in the opposite direction. 

A first point of debate raised by Newton's formulation was the status of the notion of {\em force}.  
Was Newton's second law in fact a {\em definition} of impressed force, or did it express the physical
consequence of the existence of various known and unknown forces? This was the more confusing 
as for Newton and his contemporaries {\em force} could refer to a large variety of physical quantities
related to motion, including not only the impressed force, such as contact forces involved in collisions
or central forces like the gravitational force, but among others also inertia, energy or power
\ct{jammer:1957}.

Then there was the question whether the principle of inertia --Newton's first law-- was an independent 
rule, or simply a consequence of the second, as the absence of force on a body automatically seemed 
to imply its momentum to be constant. Linked to this was the problem of how to define (inertial) mass 
as it appears in the definition of momentum. Newton defined it simply as the product of density and 
volume; this was based on the idea of a universal substance of matter, different types of matter being 
distinguished only by the number and size of empty spaces: pores or voids, in the material. But as the 
atomic nature of matter became established by the advances in chemistry in 18th an 19th century this 
picture of the structure of matter had to be abandoned and the definition of mass --as different from 
weight-- as a separate measurable property of elementary objects like atoms and molecules became 
pressing. 

And finally, did the third law imply that force was inherent to bodies, i.e.\ to matter,  or was it an 
independent immaterial physical phenomenon that became active when bodies entered each others 
sphere of influence? In particular how could a non-contact force like gravitation work instantaneously 
through empty space, and how did bodies become aware of the presence of other bodies located 
elsewhere? Was empty space really empty, or filled with some fine matter or aether?

Most interpreters in the 18th and 19th century eventually agreed that the notion of force was introduced 
as a substitute for the relation between the cause and the effect of interactions between bodies: the 
gravitational attraction of the sun is the cause of the elliptical orbit the earth moves through in the 
course of one year. But by itself this is just word-play. It becomes meaningful only if either one can 
pinpoint a definite material origin for the action of the force, such as the vortices of aether particles, 
proposed by Descartes, propelling the planets in their motion; or by accepting that it is just a convenient 
term to express mathematically the causal relations between the properties of bodies --locations, 
masses-- and the observed changes in their motion, devoid of any special physical interpretation. 

Not surprisingly this eventually lead some investigators to try to abandon the notion of force altogether.
Going back to pre-newtonian sources they realized that it was possible to replace the notion of force 
by that of energy, and to complement it by the principle used in the 16th century by Stevin for statics 
\ct{stevin:1586}, and in the 17th century by Huygens for dynamics \ct{huygens:1673}, that motion 
cannot be created out of nothing: the impossibility of a perpetuum mobile of the first kind. 

Such a reformulation of the laws of dynamics is possible only once the notion of energy is 
well-established. This was not the case in Newton's time, although Huygens and Leibniz had 
developed a theory of {\em vis viva} (living force), basically kinetic energy, for the description of 
motion of bodies in free fall and for pendulums, as studied by Galileo \ct{galileo:1632}; this theory 
was shown by Huygens to apply also to the collisions of hard bodies --in modern terminology: 
elastic collisions without dissipation of energy or momentum \ct{huygens:1703,vholten:2025}. 
The issue of {\em vis viva}, proportional to the square of the velocity of bodies, versus newtonian 
force deriving from momentum linear in the velocity of bodies, was hotly debated during the first 
half of the 18th century, and did not get fully resolved until mathematicians and natural philosophers 
such as Johann and Daniel Bernoulli, D'Alembert and Lagrange established proper definitions for, 
and connections between, force, work and energy. For a detailed summary of this controversy see 
Smith \ct{smith:2006}. 
\vs{2}

\np
\nit
{\bf 2.\ Inertial frames} 
\vs{1}

\nit
A short-coming of all early discussions of the principle of inertia is, that it does not state with respect 
to which observer frames the motion of force-free bodies is supposed to be rectilinear and uniform. 
Newton introduced the ideas of absolute space and time to this effect, but he did not explain how to 
determine translational motion with respect to absolute space in absolute time and how to distinguish 
this motion from motion with respect to other frames of reference. In fact, he at the same time 
acknowledged the Galileo-Huygens principle of relativity, that all observations of moving bodies 
only determine their relative motions, not their absolute ones. Again such contradictory statements 
led to much confusion and debate among his contemporaries.

The resolution of this problem was only achieved in 1885 by Ludwig Lange \ct{lange:1885}. He 
realized that the principle of inertia selected a special class of reference frames, in which bodies 
not subject to external forces indeed moved uniformly along straight lines; he also gave a receipt 
how to construct such inertial frames by taking the trajectories of three force-free bodies as reference.  

This important step established the independence of the principle of inertia from Newton's second 
law of motion. To wit, to identify the observation of accelerated motion as the effect of physical forces 
requires for its precise interpretation to be preceded by the establishment of an inertial reference frame, 
with respect to which acceleration is to be measured.

Not much later it also became the bedrock for the development of Einstein's theory of 
relativity, which basically deals with the relation between inertial frames.\footnote{For a recent 
discussion, see ref.\ct{koekoek:2022}.} That inertial frames are fundamental constructs in classical 
as well as in relativistic mechanics only reinforces the above conclusion. 
\vs{2}

\nit
{\bf 3.\ The impossibility of a perpetuum mobile of the first kind}
\vs{1}

\nit
It is not known when or where the first ideas about perpetual motion and its exploitation first arose. 
However, apparently the possibility of a perpetual motion machine performing work seems to 
have been contested from the start. The impossibility of creating motion for free out of nothing
was commonly stated in professional treatises and discussions on engineering, including for example 
the notebooks of Leonardo da Vinci \ct{davinci}. 

In 1586 Simon Stevin published a book on statics and mechanical equilibrium \ct{stevin:1586}.\footnote{
In this book he also describes an experiment he performed with his friend Jan de Groot, dropping canon 
balls of different weight from the church tower in the city of Delft; by listening to the sound of impact of
the canon balls on a wooden plank at the bottom of the tower they established that there was no
significant difference in the arrival time of different weights in free fall over the same distance.}
In this book he proved a theorem on the static equilibrium of weights on differently inclined planes 
(fig.\ 1): two different weights are in equilibrium if their ratio is equal to the ratio of the lengths of 
inclined planes. To show this, he imagined a string of identical beads hanging on a prism, the 
sides of which formed the inclined planes. He stated that closing the string below the prism 
would not change the condition for equilibrium, as it exerted the same force on both sides. The
crux of his argument now was, that a rotation of the string would not change the weights on either 
inclined plane formed by the sides of the prism; thus the conditions are unchanged when the 
string would move by turning either clockwise or anti-clockwise. Therefore if the string is at rest, 
it will stay at rest. There is no mechanism for creating spontaneous motion of the string. 
\bc
\scalebox{0.3}{\includegraphics{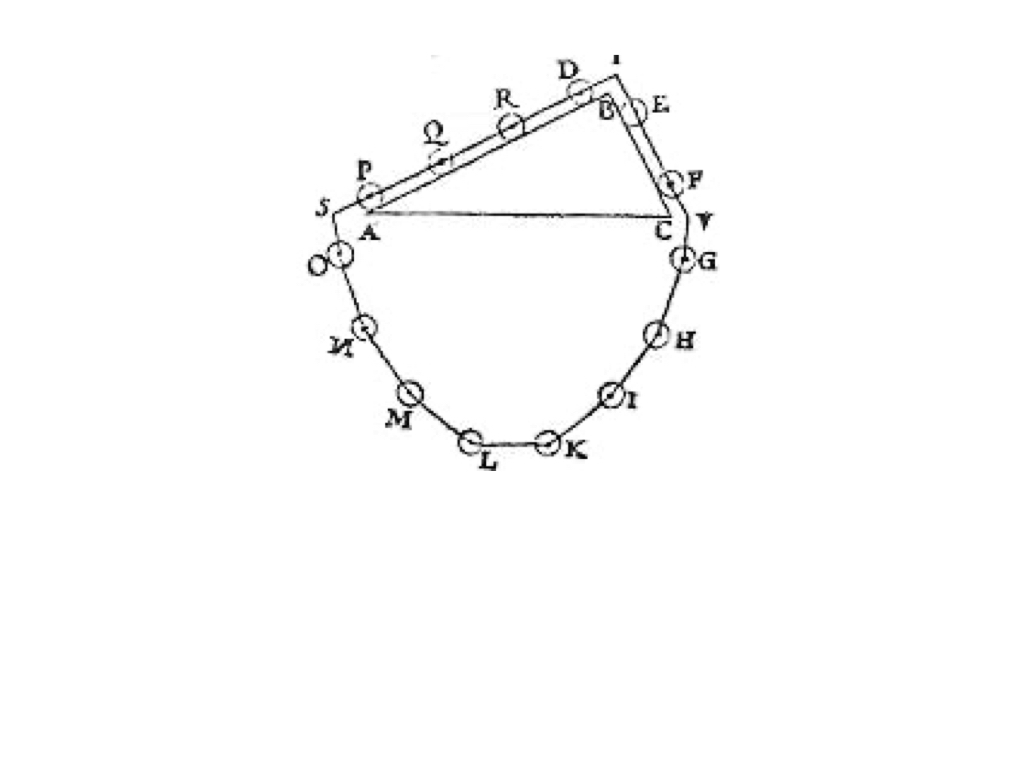}}
\vs{-14}

\footnotesize{Fig.\ 1: Stevins proof of static equilibrium of weights on inclined planes}
\ec
Somewhat later Galileo gave the first quantitative description of the motion of free fall. It states 
that in the absence of friction or air resistance a body in free fall is uniformly accelerated: it  
increases its speed $v$ linearly with time, and therefore the height $h$ it covers in descending 
from a vertical  position $z_0$ grows as the square of time; in modern notation:
\be
v = gt, \hs{2} h = z_0 - z = \frac{1}{2}\, gt^2 = \frac{1}{2g} v^2. 
\label{3.1}
\ee
He also established that the speed with which a falling body reaches its lowest point is 
precisely sufficient --in the absence of friction-- to bring it back to the same height along 
the same or another path. This is obvious in the motion of a pendulum, which starts from 
rest at a certain height, then reaches its maximum speed at the lowest point before returning 
to rest on the other side, at the same height it started from \ct{galileo:1632}. 

Huygens realized that this also was an example of the impossibility of a spontaneous 
motion machine: if it would be possible for the pendulum or the descending body to ascend
to a higher point in climbing back up, the cycle could be repeated to create an ever larger 
speed of the body at the lowest point; if in contrast motion disappeared spontaneously, not 
due to friction, then the process could simply be reversed to create motion instead of 
destroying it \ct{huygens:1673}. 

By extension of this argument Huygens, inspired by Toricelli \ct{toricelli:1644}, was able to 
show, that in the case of an arbitrary number of connected bodies falling together, if each 
body was independently reflected upward at the same velocity, the center of gravity would 
rise up to the same height it had at the start of the fall. In modern notation this rule can be 
phrased as stating that the sum of kinetic and gravitational potential energy is constant:
\be
\frac{1}{2}\, \sum_i m_i v_i^2 + \sum_i g m_i z_i = \sum_i g m_i z_{0i}
\label{3.2}
\ee
Finally, he also extended the validity of the conservation of vis viva $mv^2$ to apply 
to the collision of hard bodies: in the central collision of two bodies the sum of 
$m_1 v_1^2 + m_2 v_2^2$ is the same before and after the collision. The total 
momentum is conserved as well, meaning
\be
\bfP = m_1 \bfv_1 + m_2 \bfv_2 
\label{3.3}
\ee
does not change in elastic collisions. This he argued from first principles \ct{huygens:1669}, 
based on symmetries between the bodies and their motions involved; in fact, it also follows 
from his principle of relativity \ct{vholten:2025}. Indeed, the living force of two bodies in 
collision, described in a frame moving with arbitrary constant velocity $\bfu$ w.r.t.\ the 
original one, is 
\[
m_1 (\bfv_1 - \bfu)^2 + m_2 (\bfv_2 - \bfu)^2 = m_1 \bfv_1^2 + m_2 \bfv_2^2 
 - 2 \bfu \cdot ( m_1 \bfv_1 + m_2 \bfv_2) + (m_1 + m_2) \bfu^2
\]
If living force is conserved in both this and in the original frame, then the total momentum 
(\ref{3.3}) must be conserved as well.\footnote{It goes without saying that the total mass 
$m_1 + m_2$ is also taken to be the same during the whole process and in both frames.}

Leibniz realized that the argument had wider significance and formulated the principle of 
conservation of living force as an alternative to Newton's theory of motion \ct{leibniz:1686}. 
But the connection between the principle and Newton's laws was realized in full later, in 
the work of such successors as Lagrange, Helmholtz and Mach 
\ct{lagrange:1788, helmholtz:1847,mach:1871}. Each of these authors considered Newton's
impressed force as a mathematical expression for the mechanism of causality, and realized 
that it was connected to living force by the principle of work:
\be
\frac{1}{2}\, d (m \bfv^2) = m \bfa \cdot \bfv dt = \bfF \cdot d\bfx.
\label{3.4}
\ee 
Thus the living force of Huygens and Leibniz is changed by the work done on a body 
by an impressed force. The impossibility of creating motion out of nothing then is expressed 
mathematically by the statement that no net work is done on a body by a force when the 
body returns after moving on a closed path $\gam$ to its original position with the original 
velocity: 
\be
\frac{1}{2}\, \oint_{\gam} d(m\bfv^2) = 0 \hs{1} \Leftrightarrow \hs{1} \oint_{\gam} \bfF \cdot d\bfx = 0.
\label{3.5}
\ee
The similarity of this principle to Carnot's principle in thermodynamics was noted by both 
Helmholtz and Mach. 
\vs{2}

\nit
{\bf 4.\ Momentum and mass}
\vs{1}

\nit
We have already discussed the conservation of momentum in collisions of hard bodies. 
Moreover, in all theories of classical mechanics the conservation of mass was an implicit
or explicit additional principle: like motion mass could neither be created nor destroyed. 
This was observed with great precision in chemical processes. From Newton's formulation
of the theory of gravitational forces it was also clear that mass was the source of these 
forces, and that mass and weight were proportional. Thus there was a clear parallel 
between the conservation of mass and that of electric charge as the source of electric 
forces. 

But in contrast to electric charge mass played a second r\^{o}le in mechanics: that of the 
source of inertia. That raised the question how this inertial mass could be determined 
in the absence of gravity. It was in particular Ernst Mach who occupied himself with this 
problem and proposed a solution \ct{mach:1883}. Ironically, as a positivist striving to 
formulate physics on a purely empirical basis Mach was sceptical about the existence 
of invisible atoms. As concerns the origin of inertia he proposed to replace Newton's 
absolute space and time as the primordial inertial frame by the frame of the fixed stars, 
whose combined gravitational forces created resistance to change in motion of bodies
in our region of the universe. 

As to the question of how to determine this inertial mass, Mach employed the conservation 
of momentum. Note that in 2-body collisions the conservation of momentum is equivalent 
to Newton's law of action and reaction:
\be
\bfP = \bfp_1 + \bfp_2 = \mbox{constant}
\label{4.1}
\ee
implies that in a collision $\Del \bfP = 0$, and therefore 
\be
\Del \bfp_1 = - \Del \bfp_2.
\label{4.2}
\ee
Mach's proposal was to define the inertial mass in terms of the ratio of the observed 
change in velocities: 
\be
\frac{m_1}{m_2} = \left| \frac{\Del \bfv_2}{\Del \bfv_1} \right|,
\label{4.3}
\ee
which according to Newton's law must hold in all collisions, irrespective of the actual change
in the velocities. Of course, there remains an undetermined common factor in all inertial masses, 
fixed only by choosing a unit of mass. Such a choice then also shows up in the value of the 
gravitational constant $G$. 
\vs{2}

\nit
{\bf 5.\ Rebuilding classical mechanics}
\vs{1}

\nit
The discussions of the fundamentals of classical mechanics in the preceding paragraphs 
provide the ingredients for a reformulation of the laws of motion in such a way as to avoid 
the introduction of the concept of (impressed) force, other than as a term for a mathematical 
expression: the product of mass and acceleration, void of physical or metaphysical 
connotations. To keep it concise I will explicitly restrict this exposition to the motion of point 
masses. I do not consider the motions of rigid bodies or fluids, although I do not exclude it 
can be done. I am helped in this respect by Newton who spend much time and effort on 
showing that he could discuss the orbital motion of the moon and planets in terms of a single 
representative mass point, the center of mass.

However, before the theory is presented it is necessary to make a number of introductory 
remarks. First, by assumption the motion of a point mass is described by a curve $\bfx(t)$ 
in 3-dimensional euclidean space; that is, distances are obtained from the positive unit metric:
\be
ds^2 = d \bfx \cdot d \bfx.
\label{5,1}
\ee
The instantaneous velocity of the point mass is the tangent vector to this curve:
\be
\bfv = \frac{d\bfx}{dt}, 
\label{5.2}
\ee
and the acceleration is the second derivative
\be
\bfa = \frac{d^2 \bfx}{dt^2} = \frac{d\bfv}{dt}.
\label{5.3}
\ee
For definiteness I consider the motions of a finite set of $N$ point masses; their {\em configuration}
is the set of relative positions 
\[
C = (\bfx_1, ..., \bfx_N),
\]
with respect to an arbitrary chosen origin of co-ordinates. This set of masses is supposed to 
be {\em isolated}, in the sense that only mutual interactions are involved in any change of 
their motions; in particular there is no change in the motion of any single mass point in the 
absence of other mass points. 

This is expressed more precisely by the first law of motion, confirming the existence of 
inertial frames: \\
{\bf 1st. law of motion}: \\
{\em For any finite system of isolated masses without interaction (free masses) it is possible to  
construct a euclidean co-ordinate frame in which all point masses move uniformly on 
straight lines. }
\vs{1}

\nit
{\em Remarks}: 1.\ Point masses are free, if the removal of any one or several point masses does 
not change the motion of the remaining ones; \\
2.\ No inertial frame is unique: euclidean frames moving with a uniform constant velocity $\bfu$ 
w.r.t.\ to a given inertial frame are inertial frames as well. 
\vs{1}

The second law of motion defines inertial mass by affirming the existence of a constant total 
momentum: \\
{\bf 2nd. law of motion}: \\
{\em There exists a linear combination of the velocities of a set of isolated particles w.r.t.\ an 
inertial frame: }
\be
\bfP = m_1 \bfv_1 + ... m_N \bfv_N = \sum_{a = 1}^N m_a \bfv_a,
\label{5.4}
\ee
{\em such that during all and any motion of the particles this $\bfP$ is constant in time. }
\vs{1}

\nit
{\em Remark}: The coefficients $m_i$ in this combination define the inertial masses of the 
particles. As a rescaling of the constant $\bfP$ by a fixed number $\lb$ defines another 
constant $\lb \bfP$, the masses $m_a$ are defined uniquely only up to a common scale factor, 
which is fixed by the choice of a unit of mass. Having chosen a unit of mass, the quantity 
$\bfp_a = m_a \bfv_a$ is called the {\em momentum} of particle $a$, and $\bfP = \sum_a \bfp_a$ 
is the total momentum. The total momentum can also be interpreted as the momentum 
associated with the total mass concentrated at single point, the center of mass. Indeed, 
having a definition of inertial mass, the center of mass is defined by 
\be
\bfX = \frac{1}{M}\, \sum_a m_a \bfx_a, \hs{1} \mbox{with} \hs{1} M = \sum_a m_a. 
\label{5.5}
\ee
It follows that
\[
\bfP = M\, \frac{d\bfX}{dt}.
\]
Note that the center of mass does not depend on the choice of the mass unit. 
\vs{1}

Finally the third law of motion prohibits the existence of a perpetual motion device 
of the first kind. This is done by considering the kinetic energy of point masses: 
\be
T_a = \frac{1}{2}\, m_a \bfv_a^2 =  \frac{\bfp_a^2}{2m_i},
\label{5.6}
\ee
equivalent to Leibniz's {\em vis viva}, and the requirement that the change in the total 
kinetic energy $T = \sum_a T_a$ of an isolated system of particles when its configuration 
changes from $C_1 = C(t_1)$ to $C_2 = C(t_2)$ along a continuous path $\gam(t)$ in 
configuration space depends only on the initial and final configurations. \\
{\bf 3rd.\ law of motion}: \\
{\em Along any two continuous paths $\gam_1$ and $\gam_2$ connecting the same 
particle configurations $C_1$ and $C_2$ the change in total kinetic energy of a system 
of isolated particles is the same}:
\be
\Del T = \int_{\gam_1} dT = \int_{\gam_2} dT.
\label{5.7}
\ee

\nit
{\em Remarks}: 1.\ In particular, if the second path is taken in the opposite direction it follows 
that around a closed loop 
\be
\oint dT = \int_{\gam_1} dT - \int_{\gam_2} dT = 0. 
\label{5.8}
\ee 
2.\ As discussed before, eqn.\ (\ref{3.4}), the change in kinetic energy is equal to the work done 
on the system: 
\be
\Del T = \int_{\gam} \sum_a m_a \bfa_a \cdot d\bfx_a.
\label{5.9}
\ee
The quantity $\bfF_a = m_a \bfa_a$ may be called the (impressed) force acting on particle $a$. 
However, this is merely a definition of convenience; no measurable physical quantity needs 
to be associated with it, all information is contained in the motions: the configuration, velocities 
and accelerations of the individual particles. 

Eqs.\,(\ref{5.8}) and (\ref{5.9}), reflecting eqs.\,(\ref{3.5}), together state in mathematical terms that 
an isolated system of masses can do no net mechanical work if it is to return to a state with the 
same kinetic energy. 
\vs{2}

\np
\nit
{\bf 6.\ Comparison with Newton's laws} 
\vs{1}

\nit
To show that none of the usual results of classical mechanics is at stake, it must be 
possible to deduce all the standard consequences of Newton's laws from the ones proposed 
here. To this end it will suffice to consider a system of only two isolated particles with masses 
$m_1$ and $m_2$. 

\nit
1.\ First, the existence of an inertial frame guarantees that in the absence of interaction these 
particles will move in this frame uniformly on straight lines. Moreover, the Galileo-Huygens 
principle of relativity guarantees that there is no fundamental difference between a
state of uniform motion or rest. Whether or not interactions compel the particles to
change their state of motion is a separate issue. \\
2.\ Next, the conservation of total momentum 
\[
\bfP = \bfp_1 + \bfp_2,
\]
implies that any change in the momentum of one particle is offset by an equal but 
opposite change in that of the other particle as in eqn.\,(\ref{4.2}):
\[
\Del \bfp_1 = - \Del \bfp_2.
\]
This represents the physical content of Newton's third law. \\ 
3.\ Finally eqns.\,(\ref{5.8}) and (\ref{5.9}) state that the total change in kinetic energy of 
the particles vanishes if the system follows a closed path in configuration space. 
This implies, using the standard short-hand notation $\bfF = m \bfa$, that 
\be 
 \oint \lh  m_1 \bfa_1 \cdot \bfv_1 + m_2 \bfa_2 \cdot \bfv_2 \rh dt = 
  \oint \lh \bfF_1 \cdot d\bfx_1 + \bfF_2 \cdot d\bfx_2 \rh =0.
\label{6.1}
\ee
The general solution of this constraint is
\be
\bfF_1 = m_1 \bfa_1^{(0)} - \nb_1 V, \hs{2} \bfF_2 = m_2 \bfa_2^{(0)} - \nb_2 V,
\label{6.2a}
\ee
where the $\bfa{(0)}$ represent accelerations not performing any work as they are 
perpendicular to the velocity: $\bfa^{(0)} \cdot \bfv = 0$. This happens in particular for 
rotational motions and magnetic type interactions, such that for some vector quantity $\bfb$
\be
\bfa^{(0)} = \bfv \times \bfb.
\label{6.2b}
\ee
The gradient terms $\nb V$ are the ones responsible for changes in kinetic energy 
between different configurations $C_1$ and $C_2$:
\be
\Del T = -  \int_{\gam} \lh \nb_1 V \cdot d\bfx_1 + \nb_2 V \cdot d \bfx_2 \rh = - \Del V.
\label{6.3}
\ee
The scalar $V(\bfx_1,\bfx_2)$ defines the potential energy of the particle configuration 
such that the total energy $E$ is conserved:
\be
E = T + V = \mbox{constant}.
\label{6.4}
\ee
Note, that Newton's 3rd law then requires 
\be
\nb_1 V(\bfx_1,\bfx_2) = - \nb_2 V(\bfx_1,\bfx_2) \hs{1} \mbox{and therefore} \hs{1} 
V(\bfx_1,\bfx_2) = V(\bfx_2 - \bfx_1).
\label{6.5}
\ee
As expected, the interaction between particles therefore can depend only on their relative 
separation, not on their absolute position in the inertial frame. 
\vs{2}

\nit
{\bf 7.\ Angular momentum}
\vs{1}

\nit
Newton's real concern in the {\em Principia} was with what he called central forces, 
gravitation in particular; almost the first problem he solves is to show that Kepler's area 
law holds for centripetal acceleration, and only for this kind of acceleration. Now the area 
law is equivalent to the conservation of angular momentum. In the context of the 2-body 
system in the formulation of sects.\,5 and 6, the angular momentum is defined by
\be
\bfL = m_1 \bfx_1 \times \bfv_1 + m_2 \bfx_2 \times \bfv_2.
\label{7.1}
\ee
Provided there are no accelerations associated with magnetic type interactions $(\bfb = 0)$,
and recalling that the potential energy depends only on the separation $V(\bfx_1,\bfx_2) = 
V(\bfx_2 - \bfx_1)$, the rate of change of $\bfL$ then is 
\be
\ba{lll}
\dsp{ \frac{d\bfL}{dt} }& = &  m_1 \bfx_1 \times \bfa_1 + m_2 \bfx_2 \times \bfa_2 \\
 & & \\
 & = & \dsp{ - \bfx_1 \times \nb_1 V - \bfx_2 \times \nb_2 V 
  = - \lh \bfx_2 - \bfx_1 \rh \times \nb_2 V}
\ea
\label{7.2}
\ee
Therefore this rate of change vanishes provided for some function $\lb(\bfx_1,\bfx_2)$ of
the 2-particle configuration
\[
\nb_2 V = (\bfx_2 - \bfx_1)\, \lb(\bfx_1,  \bfx_2).
\]
Now as
\[
\bfx_2 - \bfx_1 = \frac{1}{2}\, \nb_2 (\bfx_2 - \bfx_1)^2 = |\bfx_2 - \bfx_1| \nb_2 |\bfx_2 - \bfx_1|,
\]
it follows that $V$ and $\lb$ themselves must be functions of the absolute distance only: 
\[
V(\bfx_2 - \bfx_1) = V(|\bfx_2 - \bfx_1|) \hs{1} \Leftrightarrow \hs{1}
\nb_2 V = \frac{\bfx_2 - \bfx_1}{|\bfx_2 - \bfx_1|}\, V'(|\bfx_2 - \bfx_1|), 
\]
the prime denoting a derivative w.r.t.\ the single variable $|\bfx_2 - \bfx_1|$. The 
acceleration is therefore along the line between the bodies and depends on their 
absolute mutual distance only.
\vs{2}

\nit
{\bf 8.\ $N$-body systems}
\vs{1}

\nit
The laws of motion formulated in sect.\ 5 are not restricted to isolated 2-particle systems. 
In an inertial frame by fiat all free particles move on straight lines, and curvilinear motion 
derives from mutual interactions. 

The conservation of total momentum implies, that of the $N$ individual particle momenta 
only $N-1$ can vary independently:
\be
\Del \bfp_N = - \sum_{a = 1}^{N-1} \Del \bfp_a.
\label{8.1}
\ee
This is the more general form of Newton's law of action and reaction. 

Recalling eqn.\,(\ref{5.9}) the third law for $N$ particles with configuration 
$C = (\bfx_1, ..., \bfx_N)$ now implies the existence of a potential energy 
$V[C] = V(\bfx_1,...,\bfx_N)$ such that if the system evolves from configuration $C_1$
to $C_2$ along any continuous path $\gam$
\be
\Del T = \int_{\gam}  \sum_{a =1}^N m_a \bfa_a \cdot d\bfx_a  = - \int_{\gam}  \sum_{a=1}^N 
 \nb_{x_a} V \cdot d\bfx_a  = - V[C_2] + V[C_1] = - \Del V.
\label{8.2}
\ee
Again there can be additional rotation-related or magnetic type accelerations $\bfa^{(0)}$ 
not performing any work and not changing the energy as $\bfa^{(0)} \cdot \bfv = 0$ 
identically. For the motions driven by potential energy however, it follows from conservation 
of total momentum $\bfP$ that 
\be
d\bfP = \sum_{a=1}^N d \bfp_a = - \sum_{a=1}^N \nb_{x_a} V dt = 0.
\label{8.3}
\ee
This implies that $V(\bfx_1,...,\bfx_N)$ depends only on $N-1$ combinations of 
the particle positions.

Indeed eqn.\,(\ref{8.3}) implies that the center of mass always moves uniformly on a 
straight line:
\[
\bfX = \frac{1}{M}\, \sum_{a = 1}^N m_a \bfx_a = \bfu\, t, \hs{2} \bfu = \frac{\bfP}{M}.
\]
This allows to shift the inertial frame, changing all position co-ordinates by the same 
amount: 
\[
\bfxi_a = \bfx_a - \bfu t,
\]
whilst all velocities $\bfv_i$ are shifted by the same constant amount $\bfu$; the new 
frame then is again an inertial frame and all laws of motion apply in the new frame just 
as in the old one. But in the new frame the center of mass is at rest in the origin:
\be
\sum_{a=1}^N m_a \bfxi_a = 0.
\label{8.4}
\ee
The configuration space can now be parametrized equivalently by the new co-ordinates: 
$C = (\bfxi_1,...,\bfxi_N)$, and the potential energy $V[C]$, being a function on the
configuration space, its dependence on the co-ordinates $\bfxi_a$ is restricted by the 
condition (\ref{8.4}), which allows independent variations of only $N-1$ particle positions
in agreement with eqn.\,(\ref{8.3}). For the independent arguments of the potential energy 
one can take for example the separations $\bfer_a = \bfx_{a+1} - \bfx_a = \bfxi_{a+1} - \bfxi_a$, 
with $a = 1,...,N-1$. Then the potential energy can be written as a function of these 
relative positions: $V(\bfx_1,...,\bfx_N) = V(\bfer_1,...,\bfer_{N-1})$. This is the 
appropriate generalization of the 2-particle result (\ref{6.5}). Of course, as the labeling 
of the point masses $m_i$ is arbitrary, it is arbitrary which $N-1$ independent particle 
separations are used to parametrize completely the potential energy $V[C]$.

It follows, that any interaction depending only on relative separations $\bfx_a - \bfx_b$ 
is allowed, as (supposing $a > b$)
\[
\bfx_a - \bfx_b = \sum_{c=b}^{a-1} \bfer_c.
\]
A contribution to the potential energy which can be written as a function of such 
separations therefore automatically satisfies the constraint (\ref{8.3}). One of the simplest 
interactions of this form is the combination of two-particle terms:
\be
V = \sum_{a=1}^{N-1} \sum_{b = a+1}^{N} V_{ab}(\bfx_b - \bfx_a),
\label{8.5}
\ee
But although this superposition of two-body interactions is often of practical use, it is 
by no means the only possible $N$-body interaction. 
\vs{2}

\nit
{\bf 9. Special Relativity}
\vs{1}

\nit
The classical theory of motion presented here can be adapted to apply to point particles in the 
framework of special relativity as well. As mentioned in section 2, inertial frames play the same 
role in a relativistic context as in the classical theory: in an inertial frame non-interacting particles 
describe straight world lines. The first fundamental principle of relativistic particle dynamics 
therefore again is the existing of inertial frames. However, what is changed is the relation 
between inertial frames: the Lorentz transformations relating two frames mix space and time 
co-ordinates, rather than being restricted to spatial rotations and translations in space and time 
\ct{koekoek:2022}. They leave the proper time interval on a particle world-line $X^{\mu}(\tau)$ 
invariant; in my conventions:
\be
c^2 d\tau^2 = - \eta_{\mu\nu}\, dX^{\mu} dX^{\nu} = d X^{0\,2} - d \bfX^2
 = c^2 dt^2 \lh 1 - \frac{\bfv^2}{c^2} \rh,
\label{9.0}
\ee
with $\bfv = d{\bfX}/dt$ the velocity in the inertial-frame co-ordinates. 

In the relativistic context the motion of non-interacting particles is characterized by their kinematic 
4-momentum 
\be
p^{\mu} = m \dot{X}^{\mu},
\label{9r.1}
\ee
the overdot denoting a derivative with respect to proper time $\tau$. By the definition of 
proper time
\be
p_{\mu} p^{\mu}  = - m^2 c^2.
\label{9r.2}
\ee
In the frame co-ordinates the time- and space components are 
\be
p^0 = \frac{mc}{\sqrt{1 - \bfv^2/c^2}}, \hs{2} \bfp = \frac{m \bfv}{\sqrt{1 - \bfv^2/c^2}}.
\label{9r.3}
\ee
For convenience in the following we use relativistic units in which $c = 1$. Thus velocity is 
dimensionless and momentum components have the dimension of mass. The time-component 
of four-momentum can then be identified as the particle energy: $E = p^0$, and the relations 
between particle energy and three-momentum $\bfp$ are
\be
E^2 = m^2 + \bfp^2, \hs{2} \bfv = \frac{\bfp}{E}. 
\label{9.3}
\ee
It follows that any changes in time are constrained by 
\be
\frac{dE}{dt} = \frac{\bfp}{E} \cdot \frac{d\bfp}{dt} = \bfv \cdot \frac{d\bfp}{dt}.
\label{9r.4}
\ee
The crucial difference with non-relativistic particles is, that relativity and Lorentz invariance
do not allow instantaneous action at a distance. Interactions proceed necessarily through 
mediating relativistic fields propagating at finite speeds. One can however consider the 
situation in which interactions are {\em local} in the sense that all mediating fields tend to 
vanish at large distances. In that case the particles behave asymptotically like free particles 
when separated after a large time by large distances. In analogy to the non-relativistic case 
we therefore take as a fundamental Lorentz-covariant principle that in the case of $N$ locally 
interacting particles the total asymptotic four-momentum of the particles is conserved:
\be
P^{\mu} = \sum_{a=1}^N p_a^{\mu} = \mbox{constant}.
\label{9.4}
\ee
Then the ratios of the inertial masses follow as before directly from this conservation law; 
in particular for two-body scattering in the large-time limit
\be
\Del p_1^{\mu} + \Del p_2^{\mu} \rightarrow 0,
\label{9.5}
\ee
allowing to determine $m_1/m_2$ from the changes in the velocity $\Del \bfv_{1,2}$. 
As will become clear in the following, limitations to the applicability of this conservation 
law actually come from the dynamics of the mediating field itself. 

In general the dynamics of particles is changed however, because the fields of which  
they carry charges contribute to the energy-momentum 4-vector. As an example I here 
concentrate on the case of a relativistic mediating scalar field $\vf(x)$. For particles 
with scalar charge $g$ the full canonical 4-momentum is 
\be
p^{\mu} = \lh m + g\, \vf(X) \rh \dot{X}^{\mu},
\label{9r.5}
\ee
which reduces to the free kinematical 4-momentum in regions where $\vf(x) = 0$ 
and is consistent with (\ref{9.4}) if the field vanishes asymptotically. In other regions 
the scalar field acts like a local contribution to the inertial mass.\footnote{In fact, $m$
can be absorbed in a shift of the scalar field, as in the Brout-Englert-Higgs scenario;
in this paper I only consider a scalar field with minimal energy at $\vf = 0$ \ct{englert:1964,
higgs:1964,jwvh:1997}.} In particular the identities (\ref{9r.3}, \ref{9.3}) now generalize to 
\be 
E = \frac{m + g\vf}{\sqrt{1 - \bfv^2}}, \hs{2} \bfp = \frac{(m+ g\vf) \bfv}{\sqrt{1 - \bfv^2}},
\label{9.r6}
\ee
as a result of which
\be
E^2 = \bfp^2 + \lh m + g \vf \rh^2, \hs{2} \bfv = \frac{\bfp}{E}.
\label{9r.7}
\ee
The dynamics of the field needs to be based on some Lorentz-covariant principle relating 
its rates of change with its energy density and interactions. For the scalar field $\vf(\bfx,t)$ 
the propagation law is taken to be the inhomogeneous Klein-Gordon equation
\be
\lh - \Box + \mu^2 \rh \vf(x) = \sum_a g_a\, \int_{\gam} d\tau\, \del^4\lh x - X_a(\tau) \rh =
  \sum_a g_a\, \sqrt{1 - \bfv_a^2}\, \del^3 \lh \bfx - \bfX_a(t) \rh,
\label{9.6}
\ee
where the particles moving along the collection of world lines $\gam = \{X_a(t)\}$ are 
sources for the field with strengths $g_a$. The parameter $\mu = 1/\lb_C$ is the inverse 
Compton wave length of the field. In the asymptotic limit far from the sources $\mu$ 
determines the range of the scalar field as for large $r = |\bfx|$ it follows that 
$\vf(\bfx,t) \sim e^{-\mu r}$. 

For the total energy of particles and field I take
\be
E_{tot} = \sum_a \sqrt{\bfp_a^2 + \lh m_a + g_a \vf(X_a) \rh^2} + 
 \frac{1}{2}\, \int d^3x \left[ (\der_t \vf)^2 + (\nb \vf)^2 + \mu^2 \vf^2 \right],
\label{9r.8}
\ee
with $\vf(x)$ subject to the constraint (\ref{9.6}). Any change in time is then determined by 
\be
\ba{lll}
\dsp{ \frac{dE_{tot}}{dt} }& = & \dsp{ \sum_a  \left[ \bfv_a \cdot \frac{d\bfp_a}{dt} + 
 g_a \sqrt{1 - \bfv_a^2}\, \frac{d\vf(X_a)}{dt} \right] }\\
 & & \\
 & & \dsp{ +\, \int d^3x \left[ \der_t \vf \lh - \Box + \mu^2 \rh \vf +\nb \cdot \lh \der_t \vf \nb \vf \rh \right] }\\
 & & \\ 
 & = & \dsp{ \sum_a \left[ \bfv_a \cdot \frac{d\bfp_a}{dt} +  g_a \sqrt{1 - \bfv_a^2}\, \bfv_a \cdot \nb_{X_a} \vf \right] + \oint_S d^2\sg\, \bfn \cdot \nb \vf\, \der_t \vf.  }
\ea
\label{9r.9}
\ee
The last term represents radiative scalar energy flowing through a large closed surface $S$ 
far from the sources, with $\bfn$ the outward unit vector on the surface element $d^2\sg$. 
The expression for the energy of the scalar field is seen to be justified, as in the absence of 
sources (i.e., particles) and without radiation the energy of asymptotically vanishing 
fields is conserved by itself. 

As long as the surface term can be neglected (i.e., for relatively short times and/or for $\mu > 0$
and $S$ large enough) it follows that the energy of the combined system of field and particles 
is conserved as well, provided
\be
\frac{dE_{tot}}{dt} = 0 \hs{1} \Leftrightarrow \hs{1} \frac{1}{\sqrt{1 - \bfv^2}}\, \frac{d\bfp_a}{dt} 
 = - g_a \nb \vf(X_a). 
\label{9.11}
\ee
These results are consistent with the relativistic particle equation of motion
\be
\frac{dp^{\mu}}{d\tau} = \frac{1}{\sqrt{1 - \bfv^2}}\, \frac{dp^{\mu}}{dt} = 
 - g \der_{\mu} \vf(X), 
\label{9.12}
\ee
which characterizes the extremal points of the world-line reparametrization invariant action 
\[
S[X] = - \int_{\gam} d\tau \lh m + g \vf(X) \rh \sqrt{ - \dot{X} \cdot \dot{X}}.
\]

Eqn.\,(\ref{9r.9}) also points out the limitations of relativistic point-particle theories. In the 
actual world point particles never form a truly isolated system: they interact through 
retarded rather than instantaneous mediating fields --such as the above scalar potentials-- 
and their accelerations create {\em dynamical} fields which carry away energy in the form 
of radiation. This is of course familiar from the theory of electrically charged particles, losing 
energy predominantly in the form of electromagnetic waves \ct{maxwell:1865,hertz:1893}, 
but also for electrically neutral particles emitting gravitational radiation 
\ct{einstein:1918, taylor:1989}. As all particle interactions are --as far as we know-- mediated 
by dynamical fields, inescapably any classical system of a finite number of point-like particles 
eventually loses energy and motion to some kind of field energy. However, as this loss is 
irreversible the process cannot be inverted to become a perpetual motion source. Obviously 
the conservation of the total four-momentum of the particles as in eq.\,(\ref{9.4}) then is no 
longer strictly applicable. 

Nevertheless we obviously know that there exist stable, non-radiating bound systems of 
electrically charged particles like atoms and molecules. At this level of the structure of 
matter classical physics, relativistic or not, is no longer applicable and only quantum 
physics can provide accurate descriptions of dynamical processes.
\vs{3}

\nit
{\bf Acknowledgement} \\
I am indebted to prof.\ Richard Kerner (Sorbonne Univ.), Dr.\ Gideon Koekoek 
(Maastricht Univ.) and Dr. Andrea Reichenberger (TU M\"{u}nchen) for a  
preliminary reading of the manuscript.

\np

\end{document}